\documentclass[prb,twocolumn,showpacs]{revtex4}     
\usepackage{epsfig,amsmath} 

\begin{document}

\title{Magnetoconductivity of Dirac fermions in graphene under charged impurity scatterings} 

\author{Xin-Zhong Yan$^{1,2}$ and C. S. Ting$^2$}
\affiliation{$^{1}$Institute of Physics, Chinese Academy of Sciences, P.O. Box 603, 
Beijing 100190, China\\
$^{2}$Texas Center for Superconductivity, University of Houston, Houston, Texas 77204, USA}
 
\date{\today}
 
\begin{abstract}
On the basis of self-consistent Born approximation, we solve the Bethe-Salpeter matrix equations for Cooperon propagator of the Dirac fermions in graphene under the charged impurity scatterings and a weak external magnetic field. In the absence of the magnetic field, the quantum interference effect in the electric conductivity from the contribution of Cooperon propagator will be studied and possible weak localization in the system is discussed in terms of the sample length and temperature. The magnetoconductivity stemming from the quantum interference effect is calculated, and the obtained results are in good agreement with experimental measurements.
\end{abstract}


\maketitle

\section{Introduction}

The study of electronic transport of Dirac fermions in graphene is still one of the focuses in condensed matter physics. Many theoretical works have been carried out since the experiments of graphene were realized.\cite{Novoselov,Geim,Zhang} Most of these studies (including the earlier ones in advance of the experiment)\cite{Ando,Suzuura,McCann,Khveshchenko,Aleiner,Peres,Ostrovsky} are based on the model of zero-range scatters or constant impurity scattering potentials in graphene. However, it has been found that the charged impurities with screened Coulomb potentials\cite{Nomura,Hwang,Yan} are responsible for the observed carrier density dependence of the electric conductivity of graphene.\cite{Geim} Therefore, a satisfactory transport theory for Dirac ferions in graphene should take into account the charged impurity-electron scatterings.

Recently, we have developed a formalism for the electric conductivity of Dirac fermions under the charged impurity scatterings within the self-consistent Born approximation (SCBA).\cite{Yan} The formalism has been extended to study the Hall conductivity of graphene.\cite{Yan3} Our results show that the experimentally measured electric conductivity and the inverse Hall coefficient can both be successfully explained in terms of the carrier scatterings off charged impurities. Based on the same scattering mechanism, we have studied the weak localization (WL) problem in graphene\cite{Yan2} by considering only the conventional maximum crossing diagram.\cite{Suzuura} In that work, the reason why the WL in graphene has not been detected in the experiments was given. In the present paper, we go beyond our previous approach\cite{Yan2} by including two additional diagrams as suggested in Ref. \onlinecite{McCann}, and reexamine the WL effect in graphene taking into account realistic situations of experiments. The major part of our effort is to solve the Cooperon propagator under the finite-range scatters, which is quite different from the one under the zero-range scatterings as studied in the existing works.\cite{Ando,Suzuura,McCann,Khveshchenko,Aleiner,Peres,Ostrovsky}. The detailed formalism for the Cooperon propagator is presented. With the Cooperon propagator obtained in a weak magnetic field, we calculate the magnetoconductivity of graphene and compare the results with recent experimental measurements.\cite{Ki,Tikhonenko} 

The electron system in graphene at low energy excitations can be viewed as massless Dirac fermions \cite{Wallace,Ando,Castro,McCann1} as being confirmed by recent experiments.\cite{Geim,Zhang} Using the Pauli matrices $\sigma$'s and $\tau$'s to coordinate the electrons in the two sublattices ($a$ and $b$) of the honeycomb lattice and two valleys (1 and 2) in the first Brillouin zone, respectively, and suppressing the spin indices for briefness, the Hamiltonian of the system is given by
\begin{equation}
H = \sum_{k}\psi^{\dagger}_{k}v\vec
 k\cdot\vec\sigma\tau_z\psi_{k}+\frac{1}{V}\sum_{kq}\psi^{\dagger}_{k-q}V_i(q)\psi_{k} \label{H}
\end{equation}
where $\psi^{\dagger}_{k}=(c^{\dagger}_{ka1},c^{\dagger}_{kb1},c^{\dagger}_{kb2},c^{\dagger}_{ka2})$ is the fermion operator, the momentum $k$ is measured from the center of each valley, $v$ ($\sim$ 5.86 eV\AA) is the velocity of electrons, $V$ is the volume of system, and $V_i(q)$ is the finite-range impurity potential. From our previous result,\cite{Yan1} the cutoff for $k$ summation is about $k_c \approx \pi/3a$ (with $a$ the lattice constant), within which the electrons can be regarded as Dirac particles. For charged scatters, $V_i(q)$ is given by,    
\begin{equation}
V_i(q) = 
\begin{pmatrix}
n_i(-q)v_0(q)\sigma_0& n_i(Q-q)v_1\sigma_1 \\
n_i(-Q-q)v_1\sigma_1& n_i(-q)v_0(q)\sigma_0 
\end{pmatrix}\label{vi}
\end{equation}
where $n_i(-q)$ is the Fourier component of the impurity density, $v_0(q)$ and $v_1$ are respectively the intravalley and intervalley impurity scattering potentials, and $Q$ is a vector from the center of valley 2 to that of the valley 1 [Fig. \ref{fig1}(a)]. In Appendix, we detail the discussion on this impurity potential. Here, all the momenta are understood as vectors. We will adopt the charged impurity potential as the Thomas-Fermi (TF) type, $v_0(q)=2\pi e^2\exp(-qR_i)/ (q+q_{TF})$, where $R_i$ is the distance of the impurity from the graphene layer, $q_{TF} = 4k_Fe^2/v\epsilon$ is the TF wavenumber, $k_F$ is Fermi wavenumber, and $\epsilon \sim 3$ is the effective dielectric constant. (In terms of the carrier density $n$, $k_F$ is given by $k_F = \sqrt{\pi n}$. We will use the parameter $\delta$ defined as the doped carriers per carbon atom to denote the carrier concentration. It is related to $n$ via $\delta = \sqrt{3}a^2n/4$.) The intervalley scattering potential is defined as $v_1=v_0(\bar Q)$ with $\bar Q = 4\pi/3a$ as the distance between the two nearest Dirac points. In order to fit the experimental data, the impurity concentration is chosen as $n_i = 1.15\times 10^{-3}a^{-2}$.\cite{Yan} The distances $R_i$ may be within the range of a few lattice constants. Because the typical momentum transfer $q$ due to intravalley scattering is of the order of $k_F$, $qR_i << 1$ for the doping levels considered here, the electric conductivity $\sigma$ does not sensitively depend on $R_i$. Comparing to $v_0(q)$, $v_1$ is very small unless some of the impurities are close to the graphene layer. For the case of the impurities on the layer, we have $v_1 = 2\pi e^2/ (\bar Q+q_{TF})$. In our following discussion, we set $R_i = a$ and $v_1 = \alpha 2\pi e^2/ (\bar Q+q_{TF})$ with $\alpha$ an adjustable constant. 

\begin{figure} 
\centerline{\epsfig{file=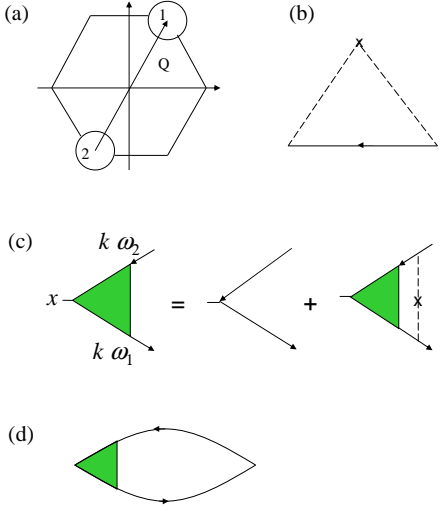,width=6.6 cm}}
\caption{(color online) (a) Brillouin zone and the two Dirac-cone valleys. (b) Self-consistent Born approximation for the self-energy. (c) Current vertex with impurity insertions. (d) Current-current correlation function. The solid line with arrow is the Green function. The dashed line is the effective impurity potential.}\label{fig1}
\end{figure} \label{fig1}

To theoretically study the electronic transport in graphene, we need to start with the Green function theory.\cite{Gorkov,Lee1,Fradkin,Altshuler,Hikami} We here briefly review the main results for the SCBA and the current vertex correction. Under the SCBA [Fig. \ref{fig1}(b)], the Green function $G(k,\omega)=[\omega+\mu-v\vec
 k\cdot\vec\sigma\tau_z-\Sigma(k,\omega)]^{-1}$ and the self-energy $\Sigma(k,\omega)=\Sigma_0(k,\omega)+\Sigma_c(k,\omega)\hat k\cdot\vec\sigma\tau_z$ of the single particles are determined by coupled integral equations:\cite{Yan} 
\begin{eqnarray}
\Sigma_0(k,\omega) &=& \frac{n_i}{V}\sum_{k'}[v^2_0(|k- k'|)+v^2_1]g_0(k',\omega)\label{sc2}\\ 
\Sigma_c(k,\omega) &=& \frac{n_i}{V}\sum_{k'}v^2_0(|k-k'|)
g_c(k',\omega)\hat k\cdot\hat k'\label{sc3}
\end{eqnarray}
with 
\begin{eqnarray}
g_0(k,\omega) &=& \frac{\tilde\omega}{\tilde\omega^2-h_{k}^2}, \nonumber\\
g_c(k,\omega) &=& \frac{h_{k}}{\tilde\omega^2-h_{k}^2} \nonumber
\end{eqnarray}
where $\tilde\omega=\omega+\mu-\Sigma_0(k,\omega)$ with $\mu$ the chemical potential, $h_k = vk+\Sigma_c(k,\omega)$, $\hat k$ is the unit vector in $k$ direction, and the frequency $\omega$ is understood as a complex quantity with infinitesimal small imaginary part. 

According to the SCBA, the current vertex is made of the ladder diagrams as shown in Fig. \ref{fig1}(c). The current vertex $v\Gamma_x(k,\omega_1,\omega_2)$ is expanded as
\begin{equation}
\Gamma_x(k,\omega_1,\omega_2)=\sum_{j=0}^3y_j(k,\omega_1,\omega_2)A^x_j(\hat k) \label{vt}
\end{equation}
where $A^x_0(\hat k)=\tau_z\sigma_x$, $A^x_1(\hat k)=\sigma_x\vec\sigma\cdot\hat k$, $A^x_2(\hat k)=\vec\sigma\cdot\hat k\sigma_x$, $A^x_3(\hat k)=\tau_z\vec\sigma\cdot\hat k\sigma_x\vec\sigma\cdot\hat k$, and $y_j(k,\omega_1,\omega_2)$ are determined by four-coupled integral equations.\cite{Yan} The $x$-direction current-current correlation function [Fig. \ref{fig1}(d)] is obtained as
\begin{eqnarray}
P(\omega_1,\omega_2)
= \frac{8v^2}{V}\sum_{kj}L_{0j}(k,\omega_1,\omega_2)y_j(k,\omega_1,\omega_2)\nonumber
\end{eqnarray}
with 
\begin{eqnarray}
L_{ij}(k,\omega_1,\omega_2) = \int_0^{2\pi}\frac{d\phi}{8\pi}{\rm Tr}[A^{x\dagger}_i(\hat k)G(k,\omega_1)A^x_j(\hat k)G(k,\omega_2)]\nonumber
\end{eqnarray}
and $\phi$ is the angle of the vector $k$, \textit{}for $\omega$'s ($\omega_1$ and $\omega_2$) = $\omega\pm i0 \equiv \omega^{\pm}$. The detailed derivations of  
$\Gamma_x(k,\omega_1,\omega_2)$ and $P(\omega_1,\omega_2)$ can be found in Ref. \onlinecite{Yan}, and will not be repeated here.

\section{Cooperon}

A major task of theoretically studying the quantum interference effect in the electronic transport is to find out the solution to the Cooperon propagator in the presence of charged impurities. Although this problem is discussed in Ref. \onlinecite{Yan2}, the detailed derivation has not been given there. In subsection A, we solve the Cooperon propagator of Dirac fermions in the absence of external magnetic field $B$. In subsection B, we present the solution to the case with the existence of the magnetic field. 

\begin{figure} 
\centerline{\epsfig{file=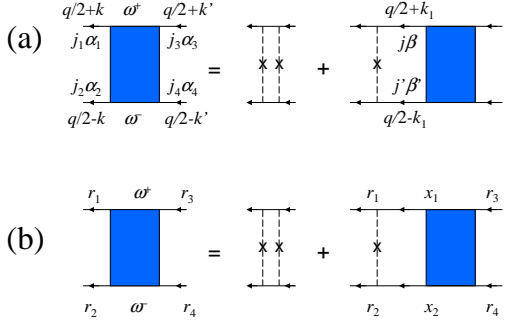,width=7.0 cm}}\label{fig2}
\caption{(color online) (a) Cooperon propagator in momentum space and (b) in real space.}
\end{figure}
 
\subsection{$B = 0$}

The Cooperon propagator $C^{j_1j_2j_3j_4}_{\alpha_1\alpha_2\alpha_3\alpha_4}(k,k',q,\omega)$ obeys the Bethe-Salpeter $16\times 16$ matrix equation represented in Fig. 2(a). Here, the superscripts $j$'s denote the valley indices, and the subscripts $\alpha$'s correspond to the sublattice indices. To explicitly write out the equation of Fig. 2(a), we here give the simpler one for $\tilde C^{j_1j_2j_3j_4}_{\alpha_1\alpha_2\alpha_3\alpha_4}(k,k',q,\omega)$ that starts from the single impurity line [the dashed line with a cross in Fig. 2(a)], using the convention $\delta^{j_1j_2}_{\alpha_1\alpha_2} = \delta_{j_1j_2}\delta_{\alpha_1\alpha_2}$ and $\bar{j}$ ($\bar\alpha$) as the conjugate valley (site) of $j$ ($\alpha$):
\begin{widetext}
\begin{eqnarray}
\tilde C^{j_1j_2j_3j_4}_{\alpha_1\alpha_2\alpha_3\alpha_4}(k,k',q,\omega) &=&
n_iv^2_0(|k-k'|)\delta^{j_1j_3}_{\alpha_1\alpha_3}\delta^{j_2j_4}_{\alpha_2\alpha_4}
+n_iv^2_1\delta^{j_1\bar j_3}_{\alpha_1\bar\alpha_3}\delta^{\bar j_2j_4}_{\bar\alpha_2\alpha_4}\delta_{j_1\bar j_2} \nonumber\\
& &+ \frac{n_i}{V}\sum_{k_1\beta\beta'}v^2_0(|k-k_1|)G^{j_1j_1}_{\alpha_1\beta}(q/2+k_1,\omega^+)G^{j_2j_2}_{\alpha_2\beta'}(q/2-k_1,\omega^-)\tilde C^{j_1j_2j_3j_4}_{\beta\beta'\alpha_3\alpha_4}(k_1,k',q,\omega) \nonumber\\
& &+ \frac{n_i}{V}\sum_{k_1\beta\beta'}v^2_1G^{\bar j_1\bar j_1}_{\bar\alpha_1\beta}(q/2+k_1,\omega^+)G^{j_1j_1}_{\bar\alpha_2\beta'}(q/2-k_1,\omega^-)\tilde C^{\bar j_1j_1j_3j_4}_{\beta\beta'\alpha_3\alpha_4}(k_1,k',q,\omega)\delta_{j_1\bar j_2}. \label{cp0}
\end{eqnarray}
\end{widetext}
The first term in the first line in right hand side of Eq. (\ref{cp0}) is due to the intravalley scatterings, while the second term comes from the intervalley scatterings. $\delta_{j_1\bar j_2}$ means that when a particle is scattered to valley $j_1$, another particle should be scattered to valley $\bar j_1$ because of the constraint for the intervalley scattering $\langle n_i(Q)n_i(Q')v_1^2\rangle/V = n_iv^2_1\delta_{Q',-Q}$. It also ensures that the total momentum (vanishing small under consideration) of the Cooperon is unchanged. The second and third lines are the processes of Cooperon propagating after the intravalley and intervalley scatterings, respectively. The equation for the Cooperon propagator $C$ is obtained by subtracting the single impurity line from $\tilde C$ in Eq. (\ref{cp0}).

The form of Eq. (\ref{cp0}) seems rather miscellaneous. It may be simplified by classifying it with good quantum number of the Cooperon. To do this, we note that the elements of the coefficient matrix of $\tilde C$ in Eq. (\ref{cp0}) are arranged according to the indices (superscripts and subscripts) of the Green functions. Since the Green function $G(k,\omega)$ are composed by the unit matrix and $\tau_z\vec\sigma\cdot\vec k$, we then look for all the operators that commute with $\tau_z\vec\sigma\cdot\vec k$. Recently, McCann {\it et al.}\cite{McCann} have introduced the operators of isospin $\Sigma$'s and pseudospin $\Lambda$'s,
\begin{eqnarray}
\Sigma_0 &= \tau_0\sigma_0,~~~\Sigma_x = \tau_z\sigma_x,~~~\Sigma_y = \tau_z\sigma_y,~~~\Sigma_z = \tau_0\sigma_z, \nonumber\\
\Lambda_0 &= \tau_0\sigma_0,~~~\Lambda_x = \tau_x\sigma_z,~~~\Lambda_y = \tau_y\sigma_z,~~~\Lambda_z = \tau_z\sigma_0.\nonumber
\end{eqnarray}
Clearly, $\Lambda$'s commute with $\Sigma$'s and $\tau_z\vec\sigma\cdot\vec k$, and are conserving operations for the Cooperon. Therefore, we transform the Cooperon from the valley-sublattice space into the isospin-pseudospin space according to McCann {\it et al.},
\begin{equation}
C^{ll'}_{ss'} =
\frac{1}{4}\sum_{\{j,\alpha\}}(M^{l}_{s})^{j_1j_2}_{\alpha_1\alpha_2}C^{j_1j_2j_3j_4}_{\alpha_1\alpha_2\alpha_3\alpha_4}(M^{l'\dagger}_{s'})^{j_4j_3}_{\alpha_4\alpha_3} \label{vlpi}
\end{equation}
where $M^l_s = \Sigma_y\Sigma_s\Lambda_y\Lambda_l$. We will hereafter occasionally use the indices 0,1,2,3 or 0,x,y,z to label $l$ and $s$. In the isospin-pseudospin space, the single impurity line is given by $W^{ll'}_{ss'} = W^l_{s}\delta^{ll'}_{ss'}$ with
\begin{equation}
W^l_{s}(|k-k_1|) = n_iv^2_0(|k-k_1|)+n_iv^2_1(\delta_{l0}-\delta_{lz})(-1)^s, \nonumber
\end{equation}
which is the transform of the first line in the right hand side of Eq. (\ref{cp0}). The result of second+third lines in right hand side of Eq. {\ref{cp0}) is transformed to 
\begin{equation}
\frac{1}{V}\sum_{k_1,s_1} [W^l(|k- k_1|)\hat h(k_1,q)]_{ss_1}\tilde C^{ll'}_{s_1s'}(k_1,k',q) \nonumber
\end{equation}
where $\hat h(k_1,q)$ is a matrix defined in the isospin space with the element given by 
\begin{eqnarray}
h_{ss'}( k_1, q) = {\rm Tr}[G(-k^+_1,\omega^+)\Sigma_sG(-k^-_1,\omega^-)\Sigma^{\dagger}_{s'}]/4 \label{hkq}
\end{eqnarray}
and $k^{\pm}_1 = k_1\pm q/2$. With these results, we obtain the equation for $C^{ll'}_{ss'}$,
\begin{eqnarray}
C^{ll'}_{ss'}(k,k',q) &=& \frac{1}{V}\sum_{k_1,s_1}\Pi^l_{ss_1}(k,k_1,q)[W^l_{s_1}(|k_1-k'|)\delta^{ll'}_{s_1s'}\nonumber \\
& & ~~~~~~+C^{ll'}_{s_1s'}(k_1,k',q)]  \label{bsh0}
\end{eqnarray}
where $\hat \Pi^l(k,k_1,q) = \hat W^l(|k- k_1|)\hat h(k_1,q)$. Here, the argument $\omega$ of $C^{ll'}_{ss'}$ and $\Pi^l$ has been suppressed for briefness. From Eq. (\ref{bsh0}), it is seen that the pseudospin of the Cooperon is indeed conserved during the impurity scatterings, $C^{ll'}_{ss'}=C^{l}_{ss'}\delta_{ll'}$. We then need to deal with $C^l_{ss'}$. Thus, the original $16\times 16$ matrix equation is separated into four $4\times 4$ ones, each of them corresponding to a definite pseudospin $l$. In the isospin space, the Cooperon of a pseudospin $l$ is a $4\times 4$ matrix denoted as $C^l$. 

To solve Eq. (\ref{bsh0}), we use the standard method that expands $C^l$ in terms of the eigenfunctions $\Psi^l_n(k,q)$ of $\Pi^l(k,k_1,q)$: 
\begin{eqnarray}
C^l(k,k',q) = \sum_nc^l_n(q)\Psi^l_n(k,q) \Psi^{l\dagger}_n(k',q), \label{exp}
\end{eqnarray}
where $c^l_n(q)$ are constants and 
\begin{eqnarray}
\frac{1}{V}\sum_{k_1}\Pi^l(k,k_1,q) \Psi_{n}(k_1,q) = \lambda^l_n(q) \Psi_{n}(k,q) \label{egn1}
\end{eqnarray}
with $\lambda^l_n(q)$ as the eigenvalue. Here, $\Psi^l_n(k,q)$ is a column vector with four components in the isospin space since $\Pi^l(k,k_1,q)$ is a $4\times 4$ matrix in this space. The constants $c^l_n(q)$ are determined by substituting Eq. (\ref{exp}) into Eq. (\ref{bsh0}). It is then seen that $c^l_n(q) \propto [1-\lambda^l_n(q)]^{-1}$. Therefore, the predominant contribution to $C^l$ comes from the state with the lowest $|1-\lambda^l_n(q)|\equiv |1-\lambda^l(q)|$ that can be vanishing small. We will here take into account only the state of the lowest $|1-\lambda^l(q)|$ for each $l$. 

Firstly, we consider the case of $l = 0$ and $q = 0$. A solution can be explicitly obtained as $\lambda^0(0) = 1$, and $\Psi^0(k,0)\equiv \Psi(k)$ with 
\begin{eqnarray}
\Psi^t(k) = [\Delta_0(k,\omega),-\Delta_c(k,\omega)\cos\phi,-\Delta_c(k,\omega)\sin\phi,0]\nonumber
\end{eqnarray}
where $\Psi^t(k)$ is the transpose of $\Psi(k)$, $\Delta_0(k,\omega)$ = Im$\Sigma_0(k,\omega^-)$, and $\Delta_c(k,\omega)$ = Im$\Sigma_c(k,\omega^-)$. The four components of $\Psi(k)$ correspond to $s = 0, x, y,z$ respectively. The solution of $\lambda^0(0) = 1$ is the most important one which gives rise to the diverging contribution to the Cooperon. One may check this result with the help of Eqs. (\ref{sc2}) and (\ref{sc3}). Actually, the above solution is just a consequence of the Ward identity (under the SCBA for the self-energy): 
\begin{equation}
{\rm Im}\Sigma(k,\omega^-) = \frac{1}{V^2}\sum_{k'}\langle V_i(k-k'){\rm Im}G(k',\omega^-)V_i(k'-k)\rangle \nonumber
\end{equation}
where $\langle\cdots\rangle$ means the average over the impurity distributions [Fig. \ref{fig1}(b)]. There are three non-vanishing components in $\Psi(k)$ because of the finite-range impurity scatterings. For the zero-range potential, only the first component of $\Psi(k)$ survives and is a constant. One then needs to solve a scalar equation instead of the matrix integral equation.

For finite but small $q$, by expanding $\hat\Pi^0(k,k',q)$ to second order in $q$ and regarding the difference from $\hat\Pi^0(k,k',0)$ as a small deviation, we then solve the problem by perturbation method. Since expanding $\hat\Pi^0(k,k',q)$ [equivalent to expanding $\hat h(k',q)$] is an elementary manipulation but tedious [because 9 elements in $\hat h(k',q)$ need to be expanded], we here just present the result. For $l \ne 0$, the difference between $\hat\Pi^l$ and $\hat\Pi^0$ comes from the intervalley scattering term in $\hat W^l$. Similarly, we can treat this difference by perturbation. For all the cases, to the first order in the perturbation, we have 
\begin{eqnarray}
\lambda^l(q) &\approx &\frac{1}{\langle\Psi|\Psi\rangle V^2}\sum_{kk'}\Psi^{\dagger}(k)\hat\Pi^l(k,k',q)\Psi(k') \nonumber\\
&\approx &\lambda^l(0) - d_lq^2, ~~~~~~~~~~~~{\rm for}~~q\to 0 \label{lmbd}
\end{eqnarray}
with
\begin{eqnarray}
\lambda^l(0) = 1-(1+\delta_{lz}-\delta_{l0})\delta\lambda \nonumber
\end{eqnarray}
\begin{equation}
\delta\lambda = n_iv_1^2\langle\Delta_0\rangle\langle\Delta_0(|g_0|^2+|g_c|^2)
+2\Delta_c{\rm Re}(g^+_0g^-_c)\rangle/\langle\Psi|\Psi\rangle 
\label{dlmbd}
\end{equation}
where $\langle\Psi|\Psi\rangle$ = $\sum_k\Psi^{\dagger}(k)\Psi(k)/V$, $\langle\cdots\rangle$ here means an average over the Fermi area, i.e., $\langle\Delta_0\rangle\equiv \sum_{k}\Delta_0(k,\omega)/V$, $g^{\pm}_{0,c} = g_{0,c}(k,\omega^{\pm})$, and the arguments $k$ and $\omega$ have been suppressed for clarity. The coefficient $d_l$ is given by
\begin{eqnarray}
d_l &=& \langle2D^l_0\Delta_0|g_c/k|^2+(D^l_0\Delta_0+D_c\Delta_c)\{|g'_0|^2+|g'_c|^2\nonumber\\
& &-{\rm Re}[g_0^+(g_0^{-}{''}+g_0^{-}{'}/k)+g_c^+(g_c^{-}{''}+g_c^{-}{'}/k)]\}\nonumber\\
& &+(D^l_0\Delta_c+D_c\Delta_0){\rm Re}
[2g_0^{+}{'}g_c^{-}{'}-g_c^+(g_0^{-}{''}+g_0^{-}{'}/k)\nonumber\\
& &-g_0^+(g_c^{-}{''}+g_c^{-}{'}/k-g_c^-/k^2)]\rangle/8\langle\Psi|\Psi\rangle \nonumber
\end{eqnarray}
where $D^l_0=\sum_{k'}W^l_0(|k-k'|)\Delta_0(k')/V$, $D_c=\sum_{k'}W^l_1(|k-k'|)\Delta_c(k')\cos\theta/V$ with $\theta$ the angle between $k$ and $k'$, and $g' = dg(k,\omega)/dk$. To the 0th order, the eigenfunction $\Psi(k)$ is unchanged. We here consider only the case of small $q$ since that is where QIC is significant.  

\begin{figure} 
\centerline{\epsfig{file=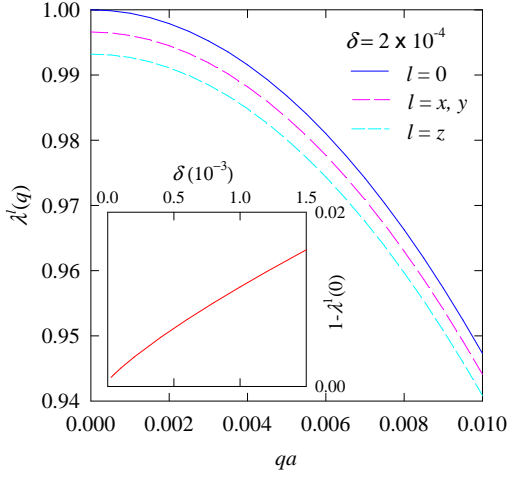,width=7. cm}}\label{fig3}
\caption{(color online) Eigenvalue $\lambda^l(q)$ as function of $q$ at electron doping concentration $\delta = 2.0\times 10^{-4}$. The inset shows $\delta\lambda$ as function of doping concentration $\delta$.}
\end{figure} 

In Fig. 3, the eigenvalues $\lambda^l(q)$ are shown as functions of $q$ at electron doping concentration $\delta = 2.0\times 10^{-4}$ and the intervalley scattering parameter $\alpha = 0.7$. The eigenvalue for $l = x,y$ is degenerated. In the limit $q \to$ 0, only $1-\lambda^0(q)$ approaches zero. As shown by Eq. (\ref{dlmbd}), the finite values $1-\lambda^1(0) = \delta\lambda$ (shown in the inset of Fig. 3) , and $1-\lambda^3(0) = 2\delta\lambda$ sensitively depend on the intervalley scattering strength $v_1$. At $v_1 = 0$, $\lambda^l(0) = 1$ is valid for all $l$. $1-\lambda^l(0)$ is larger for stronger $v_1$. On the other hand, the coefficients $d_l$ are almost the same for all the pseudospins.

The state for each $l$ so obtained is of the lowest $|1-\lambda^l(q)|$. For the lowest state, $c^l_n(q)$ is given by $c^l(q) = c_l/[1-\lambda^l(q)]$ with 
\begin{eqnarray}
c_l =\frac{1}{\langle\Psi|\Psi\rangle^2 V^3}\sum_{kk_1k'}\Psi^{\dagger}(k)\hat\Pi^l(k,k_1,0)\hat W^l(|k_1-k'|)\Psi(k') \nonumber
\end{eqnarray}
where the $q$-dependence of $c_l$ has been neglected because of the drastic behavior of $1/[1-\lambda^l(q)]$ at small $q$. The Cooperon is finally approximated as
\begin{eqnarray}
C^l(k,k',q) = c_l\Psi(k)\Psi^{\dagger}(k')/[1-\lambda^l(q)]. \label{acp}
\end{eqnarray} 
For the zero-range scatters, only the isospin-singlet $C^l_{00}(k,k',q)$ survives and is independent of $k$ and $k'$. In this case with $v_1 = 0$ and $\omega = 0$, by the one-band approximation, one obtains $\lambda^l(q) = 1-Dq^2/4\Delta_0$ where $D = v^2/2\Delta_0$ is the diffusion constant and $\Delta_0$ is the first component of $\Psi$. The resultant Cooperon is $4\Delta_0n_iv_0^2/Dq^2$ consistent with that of Ref. \onlinecite{McCann} to the order of $q^{-2}$ in $q \to 0$. 
 
\subsection{$B \ne 0$}

The Cooperon in the existence of weak magnetic field can be treated in a way parallel to Ref. \onlinecite{Altshuler}. Here, we outline the main steps. When a weak external magnetic field is applied to the system, we need to start with the description in real space because of its inhomogeneity. The kinetic part of the Hamiltonian $H_0$ is 
\begin{equation}
H_0 = \int d\vec r\tilde\psi^{\dagger}(r)v\tau_z\vec\sigma
 \cdot(\vec p + \vec A)\tilde\psi(r) \label{Hm}
\end{equation}
where $\vec p$ is the momentum operator, and $\vec A$ is the vector potential. Here, we have used the units of $e = c = 1$ (with $-e$ as an electron charge and $c$ the light velocity). With gauge transform \begin{equation}
\tilde\psi(r) = \psi(r)\exp(-i\int^r_{r_0}d\vec l \cdot\vec A), \nonumber
\end{equation}
$\vec A$ is eliminated from $H_0$. On viewing this gauge transform, for very weak magnetic field $B$, the Green function $\tilde G(r,r',\tau-\tau') = -\langle T\psi(r,\tau)\psi^{\dagger}(r',\tau')\rangle$ can be approximated as\cite{Altshuler}
\begin{equation}
\tilde G(r,r',\omega) \approx G(r-r',\omega)\exp(i\int^{r'}_{r}d\vec l\cdot \vec A), \label{gf}
\end{equation}
where $G(r-r',\omega)$ is the Green function in the absence of the magnetic field. Here, the position $r$ is understood as vector.

In real space, the Bethe-Salpeter equation for the Cooperon is diagrammatically shown in Fig. 2(b). One can then write out it explicitly and do the same transform as from Eq. (\ref{cp0}) to Eq. (\ref{bsh0}). The final matrix equation (in the isospin space) is given by 
\begin{widetext}
\begin{eqnarray}
\hat C^{l}(r_1,r_2,r_3,r_4) &=& \hat W^l(r_1-r_2)[\hat h(r_1,r_2,r_3,r_4)\hat W^l(r_3-r_4) 
+\int dx_1\int dx_2\hat h(r_1,r_2,x_1,x_2)\hat C^l(x_1,x_2,r_3,r_4)]  \label{cpr}
\end{eqnarray}
where $\hat W^l(r)$ is the real space representation of $\hat W^l(q)$, and the element of $\hat h(r_1,r_2,r_3,r_4)$ is given by
\begin{eqnarray}
h_{ss'}(r_1,r_2,r_3,r_4) = \frac{1}{4}{\rm Tr}[\tilde G(r_1,r_3,\omega^+)\Sigma_s\tilde G(r_2,r_4,\omega^-)\Sigma^{\dagger}_{s'}].\nonumber
\end{eqnarray}
Using the approximation given by Eq. (\ref{gf}), we have
\begin{eqnarray}
h_{ss'}(r_1,r_2,r_3,r_4) &\approx & h^0_{ss'}(r_1-r_3,r_2-r_4)\exp(i\int_{r_1}^{r_3}d\vec l\cdot\vec A+i\int_{r_2}^{r_4}d\vec l\cdot\vec A)   \nonumber\\
&\approx & h^0_{ss'}(R-R'+\frac{r-r'}{2},R-R'-\frac{r-r'}{2})\exp(i2\int_{R}^{R'}d\vec l\cdot\vec A) \label{hr}
\end{eqnarray}
where $R = (r_1+r_2)/2$, $r = r_1-r_2$,  $R' = (r_3+r_4)/2$, $r' = r_3-r_4$, and $h^0$ is defined in the absence of $\vec A$. The Fourier transform of $\hat h^0(R-R'+\frac{r-r'}{2},R-R'-\frac{r-r'}{2})$ is given by
\begin{eqnarray}
\hat h^0(R-R'+\frac{r-r'}{2},R-R'-\frac{r-r'}{2}) = \frac{1}{V^2}\sum_{kq}\hat h(k,q)e^{i\vec q\cdot(\vec R-\vec R')+i\vec k\cdot(\vec r-\vec r')}
\end{eqnarray}
where $h(k,q)$ is defined by Eq. (\ref{hkq}). The eigenvalue problem of Eq. (\ref{cpr}) reads
\begin{eqnarray}
\hat W^l(r_1-r_2)\int dr'_1\int dr'_2\hat h(r_1,r_2,r'_1,r'_2)\psi(r'_1,r'_2) = E^l\psi(r_1,r_2).  \label{engr}
\end{eqnarray}
where $\psi(r_1,r_2)$ is a four-component vector in the isospin space, and $E^l$ is the eigenvalue. Using the coordinates $R$ and $r$, we separate $\psi(r_1,r_2)$ as $\psi(r_1,r_2) = \Phi(R)\Psi(r)$
where $\Phi(R)$ is a scalar representing the motion of center of mass, and $\Psi(r)$ is a four-component vector meaning the relative motion of the Cooperon. Since the magnetic field is weak, only the large-scale motion of the center of mass is significantly affected; the magnetic field influence on the relative motion is negligible. Then, $\Psi(r)$ can be considered as the real space representation of $\Psi(k)$ given in Sec. II. By integrating out the relative motion, we get
\begin{eqnarray}
\frac{1}{V}\sum_{q}\int dR'\lambda^l(q)\exp[i\vec q\cdot(\vec R-\vec R')+i2\int_R^{R'}d\vec l\cdot \vec A]\Phi(R') = E^l\Phi(R)  \label{cm}
\end{eqnarray}
\end{widetext}
with $\lambda^l(q)$ defined by Eq. (\ref{lmbd}). Using $\lambda^l(q)\approx \lambda^l(0) - d_lq^2$ for small $q$, $\Phi(R') = \exp[(\vec R'-\vec R)\cdot\nabla]\Phi(R)$, and carrying out the $q$-summation and $R'$-integral, we obtain
\begin{eqnarray}
[\lambda^l(0)-d_l(\vec P+2\vec A)^2]\Phi(R) = E^l\Phi(R),  \label{cms}
\end{eqnarray}
where $\vec P = -i\nabla$ is the momentum operator (of the center of mass) of the Cooperon. Using the Landau gauge $\vec A = (0,Bx,0)$, one has
\begin{eqnarray}
E^l_n = \lambda^l(0) - 4d_lB(n+1/2), \label{LD}
\end{eqnarray}
and $\Phi_n(R)$ being the wavefunction of the corresponding Landau state. The degeneracy of each level is $g = BV/\pi$. In real space, the Cooperon is expressed as
\begin{eqnarray}
\hat C^{l}(r_1,r_2,r_3,r_4) = g\sum_n c_l\frac{\Phi_n(R)\Phi_n^{\dagger}(R')}{1-E^l_n}\Psi(r)\Psi^{\dagger}(r'). \nonumber
\end{eqnarray}

\section{Quantum interference correction and magnetoconductivity}

The magnetoconductivity measured in experiments\cite{Ki,Tikhonenko} can be shown to stem from the quantum interference correction (QIC) to the electric conductivity. We here firstly discuss the QIC in the absence of the magnetic field. For the conventional electron systems, the main contribution to the QIC is theoretically given by the maximum crossing diagrams as shown by Fig. 4(a).\cite{Fradkin,Abrahams,Lee} The works based on this diagram have also been done to understand the WL problem in graphene.\cite{Yan2,Suzuura} However, as it has been pointed out\cite{McCann} that there exist additional diagrams such as those in Fig. 4(b) which have the same order as Fig. 4(a). When the Fermi energy is not much larger than the level broadening due to the impurity scatterings, the contribution from other higher order diagrams may not be negligible.\cite{Ostrovsky} In Fig. 5, we dipict the ratio of the Fermi-level broadening $\gamma$ and the Fermi energy $E_F$ as function of the carrier concentration $\delta$ obtained for the impurity density $n_i = 1.15\times 10^{-3}a^{-2}$ and the intervalley scattering parameter $\alpha = 0.7$. At $\delta < 1.0\times 10^{-4}$, this ratio becomes larger than unity, the present approach based on the perturbation expansion my not be applicable. Our calculation for comparing with experimental measurements will be performed at $\delta > 2\times 10^{-4}$ where $\gamma/E_F < 0.4$. It is important to notice that except for the ladder diagrams given by Fig. \ref{fig1}(d) which are generated from the SCBA, any theoretical schemes including the diagrams such as in Figs. 4(a) and 4(b) are not the conserving approximation. The QIC is meaningful and qualitative only when it gives small correction to the conductivity. In this sense, the results presented below are qualitative and applicable only when the carrier density is not too small. 

\begin{figure}
\centerline{\epsfig{file=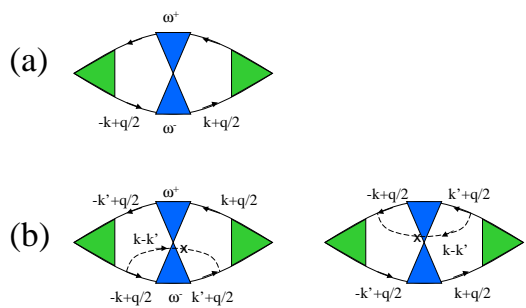,width=7.5 cm}}
\caption{(color online) (a) Maximum-crossing-diagrams correction to the conductivity. (b) Additional quantum interference correction to the conductivity.}\label{fig4}
\end{figure} 

\begin{figure} 
\centerline{\epsfig{file=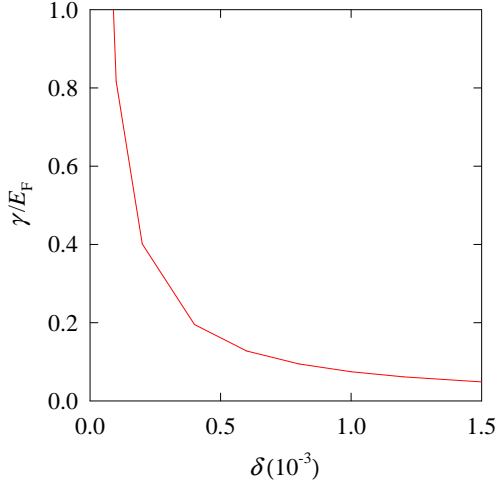,width=7. cm}}
\caption{(color online) Ratio between Fermi-level broadening $\gamma$ and the Fermi energy $E_F$ as function of carrier concentration $\delta$ calculated at impurity density $n_i = 1.15\times 10^{-3}a^{-2}$ and intervalley scattering parameter $\alpha = 0.7$.}\label{fig5}
\end{figure} 

With the Cooperon $C^l$, we firstly calculate the QIC from Fig. 4(a) to the current-current correlation function $\delta P_1(\omega^-,\omega^+)$. Because the vertex, the Green functions and the Cooperon are matrices, one needs to write out $\delta P_1(\omega^-,\omega^+)$ carefully. For doing it, we start to work in the valley-sublattice space. According to the Feynman rule, we have
\begin{widetext}
\begin{eqnarray}
\delta P_1(\omega^-,\omega^+)= \frac{2v^2}{V^2}\sum_{k q j \alpha} [V_x(-k,\omega^-,\omega^+)]^{j_4j_1}_{\alpha_4\alpha_1} C^{j_1j_2j_3j_4}_{\alpha_1\alpha_2\alpha_3\alpha_4}(-k,k,q,\omega)[V_x(k,\omega^+,\omega^-)]^{j_3j_2}_{\alpha_3\alpha_2} \label{ccc} 
\end{eqnarray}
\end{widetext}
where $V_x(k,\omega_1,\omega_2) = G(k,\omega_1)\Gamma_x(k,\omega_1,\omega_2)G(k,\omega_2)$ is the vertex connected with two Green functions. With the inverse transform of Eq. (\ref{vlpi})
\begin{equation}
C^{j_1j_2j_3j_4}_{\alpha_1\alpha_2\alpha_3\alpha_4} =
\frac{1}{4}\sum_{ll'ss'}(M^{l\dagger}_{s})^{j_2j_1}_{\alpha_2\alpha_1}C^{ll'}_{ss'}(M^{l'}_{s'})^{j_3j_4}_{\alpha_3\alpha_4}, \nonumber
\end{equation}
we obtain
\begin{eqnarray}
\delta P_1(\omega^-,\omega^+)= \frac{v^2}{2V^2}\sum_{k q l}{\rm Tr}[Z^l(k,\omega)C^l(-k,k,q)] \label{P}
\end{eqnarray}
where $Z^l(k,\omega)$ is a matrix with elements $Z^l_{ss'}$ defined as 
\begin{equation}
Z^l_{ss'}= {\rm Tr}[V_x^t(k,\omega^+,\omega^-) M^l_sV_x(-k,\omega^-,\omega^+)M^{l\ast}_{s'}]. \nonumber 
\end{equation}
There is a simple relation, $Z^l(k,\omega) = -Z^0(k,\omega)$ for $l \ne 0$, because
\begin{eqnarray}
M^l_s = M^0_s\Lambda_l,~~~~\Lambda_lM^0_s\Lambda^{\ast}_l = -M^0_s,~~~~{\rm for}~~l = x,y,z \nonumber
\end{eqnarray}
and the operator $\Lambda_l$ commutes with $G$ and $\Gamma_x$. This result means that the QIC by the pseudospin singlet ($l = 0$) is negative, while it is positive by the pseudospin triplets ($l = x,y,z$). Substituting the results given by Eqs. (\ref{lmbd}) and (\ref{acp}) into Eq. (\ref{P}) and carrying out the $q$-integral, we get 
\begin{eqnarray}
\delta P_1(\omega^-,\omega^+) = f\sum_lN_l\frac{c_l}{d_l}\ln\frac{1-\lambda^l(0)+d_lq^2_1}{1-\lambda^l(0)+d_lq^2_0},\label{P1}\\
f = \frac{v^2}{8\pi V}\sum_{k}\Psi^{\dagger}(k)Z^0(k,\omega)\Psi(-k),
\end{eqnarray}
where $N_0 = -1$, $N_{l=x,y,z} = 1$, $q_0$ and $q_1$ are the lower and upper cutoffs of the $q$-integral. 

The lower cutoff $q_0$ is given by $q_0 = {\rm max}(L^{-1}_{in},L^{-1})$ where $L_{in}$ is the length the electrons diffuse within an inelastic collision time $\tau_{in}$ and $L$ the length scale of the system.\cite{Lee} At low doping $\delta < 1.0\times10^{-3}$, $\tau_{in}$ due to the inter-electronic Coulomb interaction is estimated as $\tau_{in} \approx 0.462v/aT^2$ (where $T$ is the temperature) from the recent study of the interacting electrons in graphene using renormalized-ring-diagram approximation.\cite{Yan1} $L_{in}$ is then given by $L_{in} = (v^2\tau\tau_{in}/2)^{1/2}$ where the elastic collision time $\tau$ is determined by the non QIC-corrected conductivity $\sigma_0$, $\tau = \hbar\pi \sigma_0/vk_Fe^2$ (with $k_F$ as the Fermi wavenumber).\cite{Lee} For low carrier density, we find that $L_{in}$ is about a few microns for 4 K $< T <$ 20 K. On the other hand, the upper limit is $q_1 = L_0^{-1}$ with $L_0 = v\tau$ as the length of mean free path.

Similarly, we can obtain the contribution from diagrams in Fig. \ref{fig4}(b). Since the contribution from each of the two diagrams can be shown to be complex conjugate of each other, we then find out the formula from the first diagram, take the real part and multiply it by factor 2. The result is
\begin{widetext}
\begin{eqnarray}
\delta P_2(\omega^-,\omega^+)&=& \frac{v^2}{V^3}\sum_{kk'qlss'}{\rm Re Tr}\{\tilde C^l_{ss'}(k,k',q) V^t_x(k,\omega^+,\omega^-)M^l_{s'}G(-k,\omega^-)[V_x(-k',\omega^-,\omega^+)M^{l\ast}_sG^t(k',\omega^-)n_iv^2_0(k-k') \nonumber\\
&&+\tau_1\sigma_1V_x(-k',\omega^-,\omega^+)M^{l\ast}_sG^t(k',\omega^-)\tau_1\sigma_1n_iv^2_1]\}. \label{hk}
\end{eqnarray}
\end{widetext}
Since the maximum crossing diagrams in Fig. \ref{fig4}(b) include the single impurity line, we here have
\begin{eqnarray}
\tilde C^l_{ss'}(k,k',q)=C^l_{ss'}(k,k',q)+W^l_s(k-k')\delta_{ss'}. \label{tcp}
\end{eqnarray}
Corresponding to the approximation for $C^l_{ss'}(k,k',q)$, we also expand the matrix $W^l(k-k')$ in terms of the eigenfunctions of the peudospin-singlet Cooperon and take only the ground-state contribution. Then, the $q$-integral of $\tilde C^l_{ss'}(k,k',q)$ gives rise to a logarithm term and an additional term due to respectively $C^l$ and $W^l$ in the right hand side of Eq, (\ref{tcp}). With comparing to $\delta P_1(0^-,0^+)$, $\delta P_2(0^1,0^+)$ varies in the range of -0.3 to -0.7 $\delta P_1(0^-,0^+)$ for carrier concentration $\delta < 2\times 10^{-3}$ at low temperature. This feature is some what different from that of Ref. \onlinecite{McCann} in which the ratio is -0.5 independent of $\delta$ for zero-range impurities.  

\begin{figure} 
\vskip 5mm
\centerline{\epsfig{file=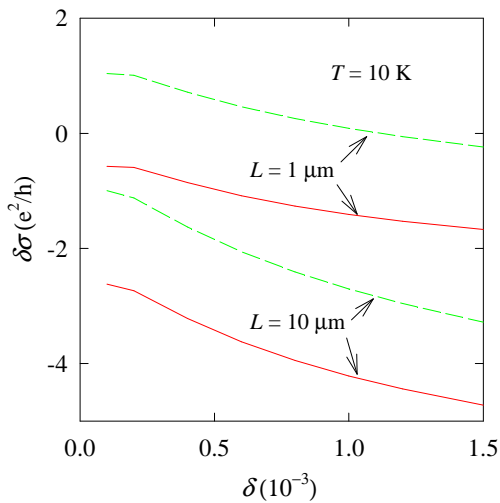,width=7. cm}}\label{fig5}
\caption{(color online) Quantum interference correction to the conductivity as function of electron doping concentration $\delta$ for the intervalley scattering parameters $\alpha = 0.7$ (solid line) and 0.5 (dashed line).}
\end{figure} 

With the current-current correlation function, one then calculates the conductivity $\sigma$ according to the Kubo formalism.\cite{Mahan} At very low temperatures, the correction to the conductivity is calculated by
$\delta\sigma = \delta P(0^-,0^+)/2\pi$. The QIC to the electric conductivity as function of electron doping concentration is shown in Fig. 6. For the intervalley scattering parameter $\alpha = 0.5$, the QIC is positive at low doping for the sample size $L = 1\mu$m, implying the system may be delocalized. While at large doping the correction becomes negative. Physically, at small electron doping, the Fermi circle and the typical momentum transfer $q$ ($\sim 2k_F$) are small and also the screening is weak, leading to stronger $v_0(q)$ than $v_1$. For very weak intervalley scattering, all $\lambda^l(0)$'s ($l \ne 0$) close to 1 (see the gap $\delta\lambda$ as function of $\delta$ in the inset of Fig. 3), the QIC from the pseudospin-triplet channels has almost the same magnitude. After one of $l \ne 0$ is canceled by the $l = 0$ channel, the net QIC is positive. This is consistent with the fact that Dirac fermions cannot be scattered to exactly the backwards direction in case of $v_1 = 0$ and the WL is absent. On the other hand, with increasing electron doping, the strength of the intervalley scatterings becomes stronger, leading to the appearance of WL in large size samples. The strength of QIC can be weakened/quenched when the sample size $< L_{in}$ (about a few microns at low temperatures) as studied in the experiment.\cite{Geim,Morozov} In large size samples, the WL effect is observable at finite carrier doping. The present result is qualitatively the same as of our previous study\cite{Yan2} where the numerical value, however, is not accurate due to a numerical error. The present model of the impurity scatterings and the results should be more realistic than the previous ones. 

We here concisely explain why pseudospin singlet Cooperon give rise to WL (negative QIC) but the pseudospin triplets result in anti-WL (positive QIC). As seen from the matrix $M^l_s$ defined below Eq. (\ref{vlpi}), the pseudospins are actually associated with the matrices $\Lambda_y\Lambda_l$, 
\begin{eqnarray}
\Lambda_y\Lambda_0 &=& \tau_2\sigma_3,~~~{\rm singlet}\nonumber\\ 
\Lambda_y\Lambda_x &=& -i\tau_3,~~~{\rm triplet~with}~l = x \nonumber\\
\Lambda_y\Lambda_y &=& 1,~~~~~~~{\rm triplet~with}~l = y \nonumber\\
\Lambda_y\Lambda_z &=& i\tau_1\sigma_3, ~~{\rm triplet~with}~l = z. \nonumber 
\end{eqnarray}
In terms of the $c$-operator, $\psi^t_{q/2-k}\Lambda_y\Lambda_l\psi_{q/2+k}$ annihilates a Cooperon of total momentum $q$ and relative momentum $k$ with pseudospin $l$. It is seen that in the pseudospin singlet state the two particles are in different valleys and the parity is odd under the exchange of the valley indices. The magnitude of the wave function of the pseudospin singlet Cooperon is large when the two particles occupy respectively the opposite momentum [defined respect to the origin of the Brillouin zone, see Fig. \ref{fig1}(a)] states of the single particles. It implies a strong backward scattering for the electrons and thereby leads to WL. On the other hand, the pseudospin triplets are even under the valley exchange. For $l = z$, though the two particles are in different valleys, the wave function of the Cooperon is small when the two particles occupy respectively the opposite momentum states. In this case, the backward scattering is weakened, resulting in the increase of the conductivity. For $l = x$ and $y$, the two particles occupy the states in the same valley and their total momentum is finite. The case corresponds to the final state of the scattered electrons is not in the backward direction, giving rise to a positive contribution to the conductivity. The anti-WL can be reduced when the backward scattering is strengthened by the intervalley scatterings.  

\begin{figure} 
\centerline{\epsfig{file=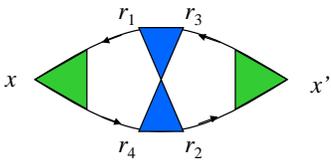,width=5. cm}}\label{fig6}
\caption{(color online) Real space representation of the quantum interference correction to the conductivity corresponding to Fig 2(a).}
\end{figure} 

The QIC to the conductivity in the presence of magnetic field is calculated according to Figs. 4(a) and 4(b). These diagrams can be redrawn in real space. For example, shown in Fig. 7 is the real space representation of Fig 4(a). To explicitly write out the expression, we note that the magnetic field effect on the vertex is negligible small. The strong dependence of $B$ in QIC comes from the Cooperon due to the small denominator $1-E^l_n$. Therefore, the vertex associated with $x'$ in Fig. 7 connected with two Green functions after integrating over $x'$ is given by
\begin{eqnarray}
V_{x'}(r_3-r_2,\omega^+,\omega^-) = \frac{1}{V}\sum_{k_1}V_{x}(k_1,\omega^+,\omega^-)e^{i\vec k_1\cdot(\vec r_3-\vec r_2)} \nonumber
\end{eqnarray}
where $V_{x}(k_1,\omega^+,\omega^-)$ has appeared above Eq. (\ref{P}). Similarly, for the vertex associated with $x$ in Fig. 7, one gets
\begin{eqnarray}
V_{x}(r_4-r_1,\omega^-,\omega^+) = \frac{1}{V}\sum_{k_2}V_{x}(k_2,\omega^-,\omega^+)e^{i\vec k_2\cdot(\vec r_4-\vec r_1)}. \nonumber
\end{eqnarray}
The QIC to the current-current correlation function from Fig. 7 is 
\begin{widetext}
\begin{eqnarray}
\delta P_1(\omega^-,\omega^+)= \frac{v^2}{2V}\sum_{l}\int dr_1\cdots\int dr_4{\rm Tr}[Z^l(r_3-r_2,r_4-r_1,\omega)C^l(r_1,r_2,r_3,r_4)] \label{Pr}
\end{eqnarray}
with $Z^l_{ss'}(r_3-r_2,r_4-r_1,\omega)= {\rm Tr}[V_{x'}^t(r_3-r_2,\omega^+,\omega^-) M^l_sV_x(r_4-r_1,\omega^-,\omega^+)M^{l\ast}_{s'}]$. Substituting the results for $V_{x'}$, $V_{x}$, and $C^l(r_1,r_2,r_3,r_4)$ into Eq. (\ref{Pr}), setting $k_{1,2} = \pm k+q/2$ and neglecting the small $q$-dependence in $V_{x}(\pm k+q/2,\omega_1,\omega_2)$, then using the coordinates $R$, $R'$, $r$, and $r'$, and integrating out the relative motions, we get
\begin{eqnarray}
\delta P_1(\omega^-,\omega^+) = \frac{4\pi gf}{V^2}\sum_{lnq}\frac{N_lc_l}{1-E^l_n}\int dR\int dR'\Phi_n(R)\Phi_n^{\dagger}(R')e^{i\vec q\cdot(\vec R'-\vec R)} 
= \frac{4\pi gf}{V}\sum_{ln}\frac{N_lc_l}{1-E^l_n}. \label{pcr}
\end{eqnarray}
\end{widetext}
By comparing this result with Eq. (\ref{P}), we see that the $q$-integral in Eq. (\ref{P}) is replaced with the summation over the Landau levels. By using the same $q$-cutoffs as in obtaining Eq. (\ref{P1}), only those states of energy levels $\epsilon^l_n =B(2n+1)$ (with $E^l_n \equiv \lambda^l(0) - 2d_l\epsilon^l_n$) in the range $(q^2_0/2, q^2_1/2)$ (in units of $a$ = $v$ = 1) need to be summed up. Similarly, we can obtain the QIC $\delta P_2$ from Fig. \ref{fig4}(b) or Eq. (\ref{hk}) in the presence of the magnetic field.

The magnetoconductivity is defined as $\Delta\sigma(B) = \sigma(B)-\sigma(0)$ where $\sigma$ is the corrected conductivity including the non-corrected one and the correction $\delta\sigma$.

Shown in Fig. 8 are the present results for the magnetoconductivity $\Delta\sigma(B)$ and comparison with experiments. For comparing with the experiments, we note that the overall magnitude of $\Delta\sigma(B)$ varies largely from sample to sample (of the same carrier density) and from experiment to experiment.\cite{Ki,Tikhonenko} The reason is that the intervalley scatterings varies in samples and in experiments due to the sample roughness or the experimental treatments. Therefore, the quantum interference effect is different. Instead to analyzing this variation, we here confine ourselves to see the magnetic field dependence of $\Delta\sigma(B)$ and therefore depict the results for the normalized magnetoconductivity. The solid, dashed, and dashed-dot lines in Fig. 8 are the present calculations for the parameters ($T, \delta, L$) = (0.12 K, $8\times 10^{-4}$, 1$\mu$m), (0.12 K, $1.5\times 10^{-3}$, 1$\mu$m), and (7 K, $2\times 10^{-4}$, 2$\mu$m), respectively. These sets of parameters correspond/close to the conditions for the two experiments: the red filled circles and green filled squares from Ref. \onlinecite{Ki}, the up and down triangles and the diamonds from Ref. \onlinecite{Tikhonenko}. The impurity density $n_i = 1.15\times 10^{-3}a^{-2}$ is the same as in our previous works which reproduce the zero-field electric conductivity\cite{Yan} and the Hall coefficient\cite{Yan3} of the experimental results.\cite{Geim} The intervalley scattering parameter is chosen as $\alpha = 0.7$. Besides this, there is no additional adjustable free parameter. Clearly, the present calculation is in good agreement with the experimental measurements.

Using the same diagrams as in Fig. 4, McCann {\it et al.} have also studied the magnetoconductivity adopting constant parameters for disorders.\cite{McCann} There are three scattering rates corresponding to three kinds of disorders in the formula for the magnetoresistivity. Here, because the charged impurities are considered as the predominant scatters, except for the impurity density (the same as before for fitting the electric conductivity and the Hall coefficient of the experimental measurements\cite{Yan3}), we have only the intervalley scattering as the parameter.

The magnetoconductivity is due to the WL effect. By writing Eq. (\ref{pcr}) as $\delta P_1 = \sum_{ln}\delta P_{ln}$ (similarly for $\delta P_2$), we see that the contribution $\delta P_{nl}(B)$ to the conductivity from each Landau level of the center of mass of the Cooperon is,
\begin{equation}
\delta P_{nl}(B) \propto \frac{B}{1-\lambda^l(0)+2d_l(2n+1)B}.
\end{equation}
For $\lambda^0(0) = 1$, $\delta P_{n0}$ is independent of $B$. But the number of Landau levels in the summation depends on $B$. Since the $q$-integral of Cooperon is replaced by the summation, the singlet pseudospin channel gives rise to a positive contribution to $\Delta\sigma(B)$. For the pseudospin triplet, $1-\lambda^l(0) > 0$, and at small $B << [1-\lambda^l(0)]/2d_l(2n+1)$, $\delta P_{nl}(B)$ increases linearly with $B$. But the contribution to $\delta\sigma(B)$ does not vanish at $B = 0$ because the Landau levels become continuum and the $q$-integral is restored giving rise to a constant. The result of the summation at $B = 0$ cancel with the corresponding term in $\sigma(0)$. Again, because of the summation over the discrete levels, the contribution to $\Delta\sigma(B)$ from the triplets is negative. Since the contributions from the pseudospin singlet and triplets vary differently as  $B$ varies, the final result for $\Delta\sigma(B)$ depends delicately on $B$.  

\begin{figure} 
\centerline{\epsfig{file=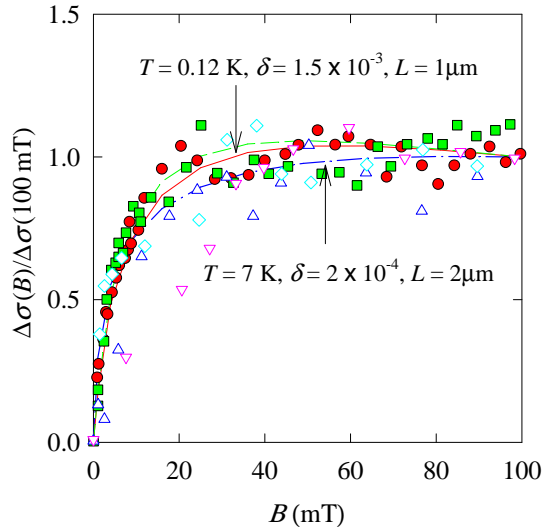,width=7.5 cm}}\label{fig7}
\caption{(color online) Normalized magnetoconductivity as function of the magnetic field $B$. The lines are the present calculations: solid, dashed, and dashed-dot lines are for the parameter sets ($T, \delta, L$) = (0.12 K, $8\times 10^{-4}$, 1$\mu$m), (0.12 K, $1.5\times 10^{-3}$, 1$\mu$m), and (7 K, $2\times 10^{-4}$, 2$\mu$m), respectively. The symbols are the experimental results: red filled circles ($\delta = 8\times 10^{-4}$ corresponding to $n = 3.15\times 10^{12}$ cm$^{-2}$) and green filled squares [$\delta = 1.5\times 10^{-3}$ ($n = 6.1\times 10^{12}$ cm$^{-2}$)], all at $T$ = 0.12 K and $L \approx 1 \mu$m, are from Ref. \onlinecite{Ki}; up triangles [$T$ = 7 K, $\delta = 2\times 10^{-4}$ ($n = 8\times 10^{11}$ cm$^{-2}$)], down triangles ($T$ = 4 K, $\delta = 2\times 10^{-4}$), and diamonds [$T$ = 0.26 K, $\delta = 2.5\times 10^{-4}$ ($n = 10^{12}$ cm$^{-2}$)] all with $L \approx 2 \mu$m are from Ref. \onlinecite{Tikhonenko}. The magnetic field is given in unit of $10^{-3}$ Tesla.}
\end{figure}

In the derivation given above, the single particle states were treated as the plane waves without taking into account the Landau quantization. We here give the estimation of its validity. In the presence of the magnetic field $B$, the Landau level of the single Dirac-fermion particle is given by\cite{McClure,Fischbeck} 
\begin{equation}
E_{n} = {\rm sgn}(n) \epsilon_0\sqrt{|n|}
\end{equation}
with $\epsilon_0 = v\sqrt{2eB/\hbar c}$. The quantity $\epsilon_0$ is the overall magnitude of the low level spacing. The above calculation of the magnetoconductivity is meaningful only when the ratio between $\epsilon_0$ and the Fermi energy $E_F = (4\pi\delta)^{1/2}v/3^{1/4}a$ is much less than unity. For the lowest doping $\delta = 2\times 10^{-4}$ and highest magnetic field $B = 0.1$ Tesla considered here, the ratio reaches about 0.1. Therefore, the present calculation shown in Fig. 8 is valid. 
 
\section{Summary}

In summary, on the basis of self-consistent Born approximation, we have solved the Bethe-Salpeter matrix equations for Cooperon propagator of the Dirac fermions in graphene under the charged impurity scatterings and weak external magnetic field. There are three non-vanishing components in the wavefunction of the Cooperon propagtor under the finite-range impurity scatterings. This feature is different from the zero-range one. The magnetoconductivity comes from the WL due to the quantum interference effect. The calculated magnetoconductivity are in good agreement with the experimental measurements.

\acknowledgments

This work was supported by a grant from the Robert A. Welch Foundation under No. E-1146, the TCSUH, the National Basic Research 973 Program of China under grant No. 2005CB623602, and NSFC under grant No. 10774171 and No. 10834011.

\vskip 5mm
\centerline {\bf APPENDIX}
\vskip 3mm

In this appendix, we discuss the impurity scattering potential. In our previous work,\cite{Yan} we have illustrated how the intravalley and intervalley scatterings $v_0(q)$ and $v_1$ in Eq. (\ref{vi}) within the SCBA are determined from the microscopic electron-impurity interactions. In our numerical calculations, $v_0(q)$ and $v_1$ are set as respectively the values of leading terms in their expansions for the charged impurities. In the derivation, the difference of the $a$ and $b$ sites in the same unit cell of graphene lattice was neglected. As long as the long-wave length scatterings are considered, such a difference is negligible. It is true for the intravalley scatterings where the long-wave length scatterings are predominant momentum transfers of the Dirac fermions. While for the intervalley scatterings, the momentum transfers are finite and more careful treatment is needed. Here, taking into account the sublattice difference, we show that only the leading terms of $v_0(q)$ and $v_1$ need to be included in the effective potentials. 

\begin{figure} 
\centerline{\epsfig{file=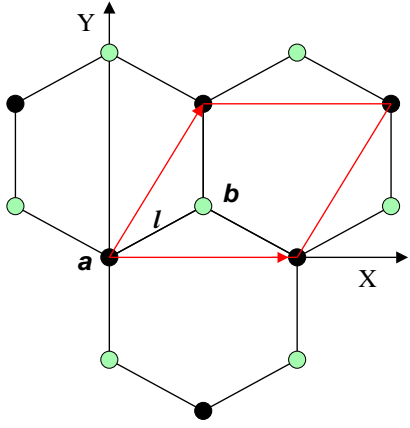,width=7. cm}}\label{fig8}
\caption{(color online) The structure of a honeycomb lattice. There are two sites, $a$ (black) and $b$ (green), in each unit cell enclosed by the red lines. {\it l} is a vector from $a$ to $b$.} 
\end{figure}

We start with the Hamiltonian for electron-impurity interactions in graphene,
\begin{eqnarray}
H_1 &=& \sum_{j\alpha}\int d\vec Rn_{\alpha}(\vec r_j)v_{\alpha i}(|\vec
 r_j-\vec R|)n_i(\vec R)  \nonumber\\
&=& \frac{1}{V}\sum_{q\alpha}n_{\alpha}(q)v_{\alpha i}(q)n_i(-q) , \label{H1}
\end{eqnarray}
where $n_{\alpha}(\vec r_j)$ is the density operator of electrons at $\alpha$ (= $a$ or $b$) site of $j$th unit cell of the honeycomb lattice (Fig. 9), $n_i(\vec R)$ is the real space density distribution of impurities, and $v_{\alpha i}(|\vec r_j-\vec R|)$ is the impurity scattering potential. The Fourier component of the electron density $n_{\alpha}(q) = \sum_kc^{\dagger}_{k-q\alpha}c_{k\alpha}$ can be written as  
\begin{eqnarray}
n_{\alpha}(q) \approx \sum_k{'} (c^{\dagger}_{k-q\alpha 1}c_{k\alpha 1}+c^{\dagger}_{k-q\alpha 2}c_{k\alpha 2}) ~~~~{\rm for}~q\sim 0, \nonumber \\
n_{\alpha}(q) \approx 
\begin{cases}
\sum\limits_k{'}c^{\dagger}_{k-q'\alpha 2}c_{k\alpha 1} ~~~~{\rm for}~q\sim Q+q'\sim Q,  \\ 
\sum\limits_k{'}c^{\dagger}_{k-q'\alpha 1}c_{k\alpha 2} ~~~~{\rm for}~q\sim -Q+q'\sim -Q, 
\end{cases}\nonumber
\end{eqnarray}
where the subscripts 1 and 2 are the valley indices [see Fig. 1(a)] and $k$ summation $\sum'$ runs over a valley since we here consider only the low energy excitations. In the main text, the $k$ summation $\sum$ means $\sum'$, and we hereafter use the simple notation $\sum$. In Eq. (\ref{H1}), the $q$ summation runs over the infinitive momentum space. By folding the whole space into the first Brillouin zone, $H_1$ can be written as 
\begin{eqnarray}
H_1 = \frac{1}{V}\sum_{kq}\psi^{\dagger}_{k-q}V_i(q)\psi_{k} , \label{H2}
\end{eqnarray}
where $k,q$ run over a valley in the Brillouin zone, and 
\begin{equation}
V_i(q) = 
\begin{pmatrix}
V_d(q)& V_o(q-Q) \\
V_o^t(q+Q)& \tilde V_d(q)
\end{pmatrix}\label{v1}
\end{equation}
with
\begin{eqnarray}
V_d(q) &=& 
\begin{pmatrix}
\phi_a(q)& 0 \\
0& \phi_b(q)
\end{pmatrix}, ~~~~
\tilde V_d(q) = 
\begin{pmatrix}
\phi_b(q)& 0 \\
0& \phi_a(q)
\end{pmatrix} \nonumber\\
V_o(q\pm Q) &=& 
\begin{pmatrix}
0& \phi^o_a(q\pm Q) \\
 \phi^o_b(q\pm Q)&0
\end{pmatrix} \nonumber\\
\phi_a(q)&=& \sum_nn_i(-q-Q_n)V(q+Q_n)\nonumber\\
\phi_b(q)&=& \sum_ne^{i(\vec q+\vec Q_n)\cdot\vec l}n_i(-q-Q_n)V(q+Q_n). \nonumber
\end{eqnarray}
Here $V(q)$ is the Fourier component of $V_{ai}(r)$, and $Q_n$ is the reciprocal lattice vector. There is a phase factor $e^{i(\vec q+\vec Q_n)\cdot\vec l}$ in the expansion of $\phi_b(q)$ because of the position difference $l$ between $a$ and $b$ sites. The expressions for $\phi^o_{a,b}(q)$ are similar to $\phi_{a,b}(q)$ except avoiding triple counting since $\phi^o_{a,b}(q\pm Q)$ imply the intervalley scatterings. [There are other two equivalent valleys for each valley indicated in Fig. \ref{fig1}(a).] For example, the leading order in $\phi^o_{a}(q-Q) \approx \phi^o_{a}(-Q)$ is $n_i(\overline Q)V(\overline Q)$ where $\overline Q$ (with $|\overline{Q}| = 4\pi/3a$, $a \sim$ 2.4 \AA~ as the lattice constant) is a  momentum difference between the nearest-neighbor Dirac points in the Brillouin zone. For this $\overline Q$, there are other two vectors of the same magnitude differ from $\overline Q$ by two reciprocal lattice vectors [see Fig. \ref{fig1}(a)], respectively. These two terms should be excluded from the summation. Since the case of $q << Q$ is under consideration, we approximate the off-diagonal potential as $V_o(q\pm Q) \approx V_o(\pm Q)$. 

We have argued\cite{Yan} that if the effective impurity scatterings are given as Eq. (\ref{vi}), then the vertex $\Gamma_x(\vec k,\omega_1,\omega_2)$ can be expanded as Eq. (\ref{vt}). [The off-diagonal parts in Eq. (\ref{vi}) are different from that in Ref. \onlinecite{Yan} where $\sigma_0$ was used instead of $\sigma_1$ because the basis was  $\psi^{\dagger}_{k}=(c^{\dagger}_{ka1},c^{\dagger}_{kb1},c^{\dagger}_{ka2},c^{\dagger}_{kb2})$ with reflected $y$-axis in valley 2. But the result for $\Gamma_x(\vec k,\omega_1,\omega_2)$ is unchanged.] Here, we want to see how $v_0(q)$ and $v_1$ in Eq. (\ref{vi}) are expected from the microscopic potentials given by Eq. (\ref{v1}). The $x$-direction current vertex $v\Gamma_x(\vec k,\omega_1,\omega_2)$ under the SCBA satisfies the following $4\times 4$ matrix integral equation,
\begin{widetext}
\begin{equation}
\Gamma_x(\vec k,\omega_1,\omega_2)=
 \tau_3\sigma_x+\frac{1}{V^2}\sum_{k'}\langle V_i(\vec k-\vec k')G(\vec k',\omega_1)\Gamma_x(\vec
 k',\omega_1,\omega_2)G(\vec k',\omega_2)V_i(\vec k'-\vec k)\rangle,\label{vt1}
\end{equation}
where $\langle\cdots\rangle$ means the average over the impurity distributions [Fig. \ref{fig1}(c)]. Notice that the product $G(\vec k,\omega_1)\Gamma_x(\vec  k,\omega_1,\omega_2)G(\vec k,\omega_2)$ can be expanded in $A^x_j(\hat k)$,
\begin{equation}
G(\vec k,\omega_1)\Gamma_x(\vec  k,\omega_1,\omega_2)G(\vec k,\omega_2) = \sum_{jj'}A^x_j(\hat k)L_{jj'}(\vec k,\omega_1,\omega_2)y_{j'}(\vec  k,\omega_1,\omega_2), \nonumber
\end{equation}
where the functions $L_{jj'}(\vec k,\omega_1,\omega_2)$ have been defined in Ref. \onlinecite{Yan}. We then need to calculate the expectations $\langle V_i(\vec k-\vec k')A^x_j(\hat k')V_i(\vec k'-\vec k)\rangle$ in Eq. (\ref{vt1}). Firstly, we calculate $\langle V_i(q)A^x_0V_i(-q)\rangle = A^x_0[\langle\phi_a(q)\phi_b(-q)\rangle-\langle\phi_a(Q)_{atc}\phi_b(-Q)_{atc}\rangle$. Using the expressions for $\phi_{a,b}(q)$, we obtain
\begin{eqnarray}
\langle\phi_a(q)\phi_b(-q)\rangle
= Vn_i\sum_nV^2(q+Q_n)e^{-i(\vec q+ \vec Q_n)\cdot \vec l} \approx Vn_iV^2(q), \nonumber
\end{eqnarray}
where the use of the fact that $\sum_{n\ne 0}|V(q+Q_n)|^2e^{-i\vec Q_n\cdot \vec l} = 0$ for $q \sim 0$ has been made. Similarly, we get
\begin{eqnarray}
\langle\phi_a(Q)_{atc}\phi_b(-Q)_{atc}\rangle 
= Vn_i\sum_n{'}V^2({\overline Q}+Q_n)e^{-i(\vec{\overline Q}+ \vec Q_n)\cdot \vec l}
\approx Vn_iV^2({\overline Q}), \nonumber
\end{eqnarray}
where $\sum_n{'}$ means avoiding triple counting, and $\vec{\overline Q}\cdot \vec l = 0$ because for $\vec l$ there is a $\vec{\overline Q}$ orthogonal to it [see Figs. 1(a) and 9]. For $A^x_j(\hat k')$ with $j \ne 0$, we note
\begin{eqnarray}
A^x_{1,2}(\hat k') = A^x_{1,2}(\hat k)(\cos\theta\pm i\sigma_z\sin\theta),~~
A^x_3(\hat k') = A^x_3(\hat k)(\cos 2\theta+i\sigma_z\sin 2\theta),
\end{eqnarray}
where $\theta$ is the angle between $\hat k$ and $\hat k'$. Since $V(\vec k'-\vec k)$ depends on $\theta$ through $\cos\theta$, the integrals of the integrands with factor $\sin\theta$ or $\sin 2\theta$ vanish. $\langle V_i(\vec k-\vec k')A^x_j(\hat k')V_i(\vec k'-\vec k)\rangle$ for $j \ne 0$ are then calculated as
\begin{eqnarray}
\langle V_i(\vec k-\vec k')A^x_{1,2}(\hat k')V_i(\vec k'-\vec k)\rangle \to A^x_{1,2}Vn_i\sum_nV^2(|\vec k-\vec k'+Q_n|)\cos\theta 
\approx A^x_{1,2}Vn_iV^2(|\vec k'-\vec k|)\cos\theta       \nonumber\\
\langle V_i(\vec k-\vec k')A^x_3(\hat k')V_i(\vec k'-\vec k)\rangle \to A^x_3Vn_i\sum_nV^2(|\vec k-\vec k'+Q_n|)e^{-i(\vec q+ \vec Q_n)\cdot \vec l}\cos 2\theta 
= A^x_3Vn_iV^2(|\vec k'-\vec k|)\cos 2\theta.       \nonumber
\end{eqnarray}
By defining $v_0^2(q) = V^2(q)$ and $v_1^2 = V^2(\overline Q)$, we obtain exactly the same equation as Eq. (11) in Ref. \onlinecite{Yan} for determining $y_j(\vec  k,\omega_1,\omega_2)$. Thus, we have proved that only the leading terms of $v_0(q)$ and $v_1$ are necessarily taken into account in the current vertex corrections. 

On the other hand, under the SCBA, the self-energy is given by
\begin{eqnarray}
\Sigma(\vec k,\omega)&=&
\frac{1}{V^2}\sum_{k'}\langle V_i(\vec k-\vec k')G(\vec k',\omega)V_i(\vec k'-\vec k)\rangle \nonumber\\
&\approx &\frac{n_i}{V}\sum_{k'}\{[\sum_nV^2(\vec k-\vec k'+Q_n)+\sum_n{'}V^2(Q-Q_n)]g_0(\vec k',\omega) + V^2(\vec k-\vec k')g_c(\vec k',\omega)\hat k\cdot\hat k'\hat k\cdot\vec\sigma\tau_z\}  \nonumber \\
&\approx &\frac{n_i}{V}\sum_{k'}\{[V^2(\vec k-\vec k')+ V^2(\overline Q)]g_0(\vec k',\omega) + V^2(\vec k-\vec k')g_c(\vec k',\omega)\hat k\cdot\hat k'\hat k\cdot\vec\sigma\tau_z\},  \nonumber 
\end{eqnarray}
\end{widetext}
which is consistent with Eqs. (\ref{sc2}) and (\ref{sc3}). Therefore, the theory is simplified with the effective potential given by Eq. (\ref{vi}) with $v_0(q)$ and $v_1$ defined above.

\end{document}